\documentclass[onecolumn,english,showpacs]{revtex4}\newcommand{\JHEPonly}[1]{}\newcommand{\PRDonly}[1]{#1}
\usepackage[dvips]{graphicx}
\usepackage{amsmath,amssymb,bm}

 \def\imo{i}

\def\imo{i}

\def\be{\begin{equation}}
\def\ee{\end{equation}}
\def\bea{\begin{eqnarray}}
\def\eea{\end{eqnarray}}

\begin{document}

\PRDonly{
\title{How to tell the shape of a wormhole by its quasinormal modes}
\author{R. A. Konoplya}\email{konoplya_roma@yahoo.com}
\affiliation{Institute of Physics and Research Centre of Theoretical Physics and Astrophysics, Faculty of Philosophy and Science, Silesian University in Opava, CZ-746 01 Opava, Czech Republic}
\affiliation{Peoples Friendship University of Russia (RUDN University), 6 Miklukho-Maklaya Street, Moscow 117198, Russian Federation}
\affiliation{Theoretical Astrophysics, Eberhard-Karls University of T\"ubingen, T\"ubingen 72076, Germany}

\begin{abstract}
Here we shall show how to reconstruct the shape function of a spherically symmetric traversable Lorenzian wormhole near its throat if one knows high frequency quasinormal modes of the wormhole. The wormhole spacetime is given by the Morris-Thorne ansatz. The solution to the inverse problem via fitting of the parameters within the WKB approach is unique for arbitrary tideless wormholes and some wormholes with non-zero tidal effects, but this is not so for arbitrary wormholes. As examples, we reproduce the near throat geometries of the Bronnikov-Ellis and tideless Morris-Thorne metrics by their quasinormal modes at high multipole numbers $\ell$.
\end{abstract}

\maketitle
}

\section{Introduction}

Recent observations of gravitational waves from compact objects \cite{TheLIGOScientific:2016src} still do not single out the geometry of the sources (though the Kerr solution is the most celebrated and probable candidate). This occurs mainly because the mass and angular momentum of the object is known with large uncertainty, so that within the current precision of measurements a non-Kerr geometry can effectively mimic the Kerr ringdown phase, allowing for not only small \cite{Cardoso-Berti}, but even relatively large deviations from Kerr geometry \cite{Konoplya:2016pmh}. Moreover, even such exotic objects, as wormholes can  mimic the ringdown phase of the Einsteinian black holes as it was first argued in \cite{Damour:2007ap}, showed in \cite{Cardoso:2016rao} and further discussed in \cite{Konoplya:2016hmd}.  The gravitational spectrum of wormholes have a wide range for interpretations \cite{Aneesh:2018hlp,Bueno:2017hyj,Maselli:2017tfq,Nandi:2016uzg,Konoplya:2010kv}. Observations in the electromagnetic spectrum also do not rule out existence of wormholes \cite{Zhou:2016koy,Tsukamoto:2017hva}.
Even though wormholes are not ruled out by current astrophysical observations, they raise a number of purely theoretical problems related to the possibility of their existence.

One of the popular approaches to construction of wormhole metrics is based on supposition of concentration of the exotic phantom matter in a thin compact region near the throat of a wormhole. At the same time no such exotic matter have been observed by now in the laboratory. If the thin shell of the exotic matter is posed at the throat and then matched with the Schwarzschild metric in such a way that the radius of the throat is very close to the Schwarzschild radius, then the Damour-Solodukhin wormhole can be constructed \cite{Damour:2007ap}. This wormhole is basically indistinguishable from the Schwarzschild black hole up to the echoes phenomena \cite{Cardoso:2016rao}. 

Here we shall consider a different situation and suppose that the Lorenzian traversable wormhole is described by the general Morris- Thorne ansatz \cite{Morris-Thorne}, so that the considered class of wormholes are symmetric relatively its throat.  This certainly is not the most general case, as it does not include the echoes effects. Neglecting the echo effects, it is reasonable to be restricted  by a class of the effective potentials which  have the form of the potential barrier, symmetric relatively its throat, so that the maximum of the effective potential coincides with the position of the throat. This occurs for a wide class of wormholes and is guaranteed, for example for all Morris- Thorne metrics whose function $g_{tt}$ is monotonic in terms of the radial coordinate $r$. Similar behavior should be expected for other massless neutral test fields in the regime of high multipole numbers $\ell$. This kind of a "good" eikonal regime behavior is usual, but not the unique for black holes and a number of wormhole solutions \cite{Morris-Thorne}.

Then, the necessary condition of viability of a wormhole's model is its stability against small perturbations of spacetime. Stability of thin-sell wormholes is tested mainly for purely radial perturbations and cannot be considered as well established, while the number of other wormholes have been proved to be unstable \cite{Bronnikov:2012ch,Cuyubamba:2018jdl,Cremona:2018wkj}. Here we shall put aside all the above existential questions (see \cite{Visser,Lobo} for reviews) and use an agnostic approach by supposing that there is a stable wormhole whose high frequency radiation obey the usual eikonal form, as, for example, it takes place for the Schwarzschild black hole \footnote{In the eikonal regime the spectrum of Schwarzschild black hole does not depend on the spin of the field under consideration \cite{Cardoso:2008bp}. However, there are a number of theories with higher curvature corrections for which the eikonal quasinormal spectrum of gravitational perturbations obeys a different law from that of test fields \cite{Konoplya:2017wot,Konoplya:2017lhs}.}.

In order to test the strong gravity regime in the vicinity of a wormhole, one needs to know the geometry of the wormhole \emph{near its throat}, while the behavior in the far region is much less interesting as the intrinsic wormhole features are mainly lost at large distance. The response of a wormhole (in the same way as of a black hole) is dominated by damped oscillations, called \emph{quasinormal modes}. Over the past decades, an extensive literature was devoted to the problem of finding of quasinormal modes for a given geometry of a compact object, be it a black hole, start or wormhole \cite{reviews}. Recently, the inverse problem - how to find the effective potential if one knows the spectrum of bound states or quasinormal modes  - was studied in \cite{Volkel:2018hwb} for a double well potential of the Damour-Solodukhin wormhole \cite{Damour:2007ap} and in \cite{Volkel:2017kfj} for ultra compact stars. However, the reconstruction of the effective potential does not lead to a single spacetime metric for black holes. In addition, for black holes there is a family of effective potentials producing the same quasinormal spectrum \cite{chandrasekhar}. Thus, because of the above reasons the ``back way''  from a quasinormal spectrum to the background metric is not unique for black holes.

Here we shall consider the following inverse problem: how to reconstruct the metric of an arbitrary traversable Lorentzian wormhole near its throat if one knows the high frequency quasinormal modes of the system. The wormhole metric is given by the general Morris-Thorne ansatz \cite{Morris-Thorne} and we shall imply that the metric functions can be expanded into Taylor series near the throat. The quasinormal spectrum will be analyzed for a test field in the regime of high multipole numbers $\ell$.

There are two important circumstances which make our analysis feasible and meaningful. The first one is the fact that the maximum of the effective potential coincides with the position of the throat, so that the near throat geometry brings much more information about a wormhole than the near horizon geometry says about a black hole. Indeed, the scattering of fields which governs such processes as quasinormal ringing, Hawking radiation, accretion of matter etc., occurs around this maximum of the effective potential, which is located at some distance from the black hole horizon.

The other aspect is related to the uniqueness of the solution to the inverse problem. As it was mentioned above, there is a family of effective potentials which produce the same quasinormal modes for black holes, so that it is impossible to extract a single effective potential from quasinormal modes. For wormholes obeying the general Morris-Thorne ansatz we have effective potentials which are symmetric relatively its throat, so that no shifting of the potential is allowed which could produce the same bound states. Thus, in the wormhole case the inverse problem for the reconstruction of the effective potential should be unique. Apparently, this is not so for the reconstruction of the metric functions themselves in the most general case, as there is an infinite set of combinations of the shape and redshift functions which can form the same effective potential. However, the usage of the WKB method enables us to reproduce the behavior of the metric near the throat for some class of wormholes with tidal effects and for arbitrary tideless wormholes.

The paper is organized as follows. In sec. II we discuss general aspects of the quasinormal mode problem for wormholes, including the master wave equation and boundary conditions. Sec. III is a quick demonstration of our method where a specific simple wormhole metric was chosen, so that only the linear terms of the Taylor expansion of the metric near the throat are taken into consideration. Sec. IV relates the general approach which includes also higher order terms. Sec. V tests the method through reconstruction of the Bronnikov-Ellis and Morris-Thorne wormhole solutions. Finally, in Sec. VI we discuss the open questions and summarize the obtained results.

\section{Quasinormal mode problem for traversable wormholes}

Static spherically symmetric Lorentzian traversable wormholes of an arbitrary shape can be modeled by a Morris-Thorne ansatz \cite{Morris-Thorne}
\begin{equation}\label{MT}
ds^2 = - e^{2 \Phi (r)} dt^2 + \frac{d r^2}{1 - \frac{b(r)}{r}} + r^2 (d \theta^2 + \sin^2 \theta d \phi^2).
\end{equation}
Here $\Phi(r)$ is the lapse function which determines the red-shift effect and tidal force of the wormhole space-time.
When $\Phi = const$ wormholes are tideless \footnote{Here we follow the terminology of \cite{Morris-Thorne} and mean that a ``tideless'' wormhole provides zero acceleration to a point particle. The tidal effects of extended bodies still remain in this case.}. A shape of a wormhole is determined by the shape function
$b(r)$.
Throat of a wormhole is situated at the minimal value of $r$, $r_{min} = b_0$. The coordinate $r$ runs from $r_{min}$ until spatial infinity $r = \infty$. In terms of the proper radial distance coordinate $d l$, given by the equation
\begin{equation}
\frac{dl}{d r} = \pm \left(1- \frac{b(r)}{r}\right)^{-1/2},
\end{equation}
there are two infinities $l= \pm \infty$ at $r= \infty$.
From the requirement of absence of singularities. $\Phi(r)$
must be finite everywhere. The requirement of asymptotic flatness gives $\Phi(r) \rightarrow 0$ as $r \rightarrow \infty $
(or $ l \rightarrow \pm \infty$). The shape function $b(r)$ must be such that $1- b(r)/r > 0$ and
$b(r)/r \rightarrow 0$ as $r \rightarrow \infty $ (or $ l \rightarrow \pm \infty$). In the throat $r = b(r)$ and thus $1- b(r)/r$ vanishes.
The metric of traversable wormholes does not have a singularity in the throat and the traveler can pass through
the wormhole during the finite time.

Here, for simplicity, we shall consider the wave equations for a test electromagnetic field,
\begin{equation}
((A_{\sigma, \alpha} - A_{\alpha, \sigma}) g^{\alpha \mu} g^{\sigma \nu}
\sqrt{-g})_{, \nu} = 0,
\end{equation}
where $A_{\mu}$ is the four-vector of the electromagnetic field.  Our analysis can be repeated for test fields of other spin in a similar way.

After making use of the metric coefficients (\ref{MT}), the perturbation equations
can be reduced to the wave-like form for the wave functions $\Psi$ (see for instance \cite{Konoplya:2006rv}):
\begin{equation}\label{sp}
\frac{d^2\Psi}{dr_*^2}+\omega^2 \Psi-\left(\frac{e^{2\Phi}\ell (\ell+1)}{r^2}\right)\Psi = 0,
\end{equation}
where the tortoise coordinate is given by
$$dr_*= \pm \frac{dr}{e^{\Phi}\sqrt{1 -\frac{b(r)}{r}}}.$$
Thus, although the explicit form of the effective potential, 
$$V = \left(\frac{e^{2\Phi}\ell (\ell+1)}{r^2}\right),$$ 
as a function of $r$ does not depend on the shape function $b(r)$ for the electromagnetic field, the resultant master wave equation and the effective potential as a function of $r^{*}$ do depend on $b(r)$ as well.

\begin{figure}
\includegraphics[width=0.6 \linewidth]{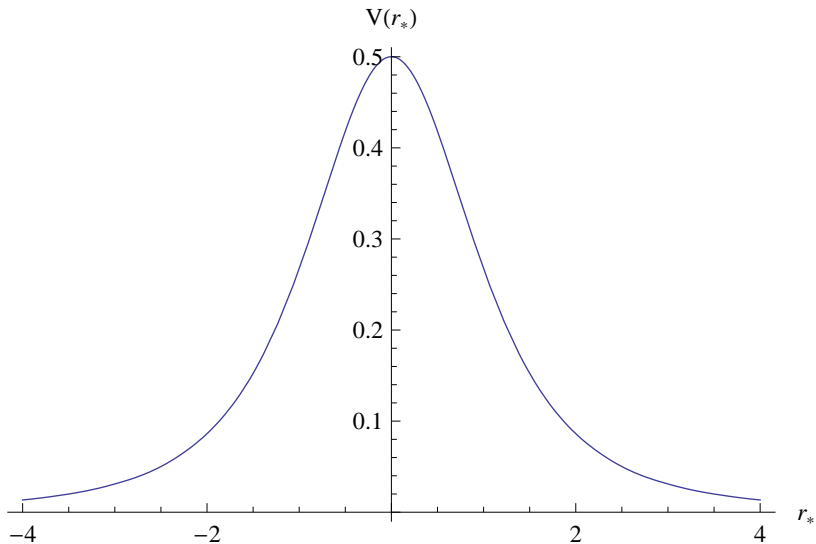}
\caption{The effective potential  for a test field as a function of the tortoise coordinate.}\label{epl0q0}
\end{figure}

In terms of the tortoise coordinate the effective potentials for metrics (\ref{MT})  have the form of the positive definite potential barriers with the peak situated at the throat of a wormhole (see Fig. \ref{epl0q0}). As it was mentioned in the introduction we neglect possible echoes from quantum corrections near the throat or from matter far from black hole, so that the double-well potential is excluded in this approach. In addition, we will analyze only the regime of high multipole numbers $\ell$, which is usually similar for fields of various spin.

The whole space lays between two ``infinities'' connecting two universes or distant regions of space. According to \cite{Konoplya:2005et}, the quasinormal modes of wormholes are solutions of the wave equation (\ref{sp}) satisfying the following boundary conditions
\begin{equation}
\Psi \sim e^{ \pm i \omega r_{*}}, \quad r_{*} \rightarrow \pm \infty.
\end{equation}
This means the pure outgoing waves at both infinities or no waves coming from either left or right infinity. This is quite natural condition if one remembers that quasinormal modes are \emph{proper} oscillations of wormholes, i.e. they represent response to the perturbation when the initial perturbation stopped acting. Alternatively, one can think of two passive observers, which do not reflect incoming radiation at both infinities.
We shall write a quasinormal mode as
\begin{equation}
\omega = \omega_{Re} + i \omega_{Im},
\end{equation}
where $\omega_{Re}$ is the real oscillation frequency and $\omega_{Im}$ is proportional to the decay rate of a given mode.

\section{Telling the metric functions near the throat by QNMs: illustration of the approach}

We will imply that the functions $b(r)$ and $\Phi(r)$ can be expanded in the Taylor series around the wormhole throat:
\begin{equation}\label{b-expansion}
b(r) = b_0 +b_1 \left(r-b_0\right) +b_2 \left(r-b_0\right)^2 +\cdots, 
\end{equation}
\begin{equation}\label{ph-expansion}
\Phi(r) = \Phi_0 + \Phi_1 \left(r-b_0\right)+ \Phi_2 \left(r-b_0\right)^2 +\cdots . 
\end{equation}
Notice that when analyzing asymptotically flat wormholes, it would be more practical to use a pre-factor providing the correct asymptotic at infinity. However, our approach does not depend on the behavior far from a wormhole, so that, for simplicity we can use the form of expansion without pre-factor.

Using the WKB formula one can find quasinormal modes in the regime of large $\ell$ in the analytical form.
This approach is based on the WKB expansion of the wave function at both infinities (the event horizon and spacial infinity) which are matched with the Taylor expansion near the peak of the effective potential. The WKB approach in this form implies existence of the two turning points and monotonic decay of the effective potentials along both infinities
\begin{equation}\label{WKB}
	\frac{i Q_{0}}{\sqrt{2 Q_{0}''}} - \sum_{i=2}^{i=p}
		\Lambda_{i} = n+\frac{1}{2},\qquad n=0,1,2\ldots,
\end{equation}
where the correction terms $\Lambda_{i}$ were obtained in \cite{WKBorder} for different orders.  Here $Q_{0}^{i}$ means the i-th derivative of $Q = \omega^2 - V$ at its maximum with respect to the tortoise coordinate $r^\star$, and $n$ labels the overtones.

Here  we shall consider the electromagnetic field, because the effective potential is simpler in that case and the final formulas are more concise. However, the privilege of usage of the WKB approach is that in the eikonal regime $\ell \rightarrow \infty$ the WKB series converges asymptotically and is accurate. In most cases the spectrum of perturbations in this regime is qualitatively the same independently on the spin of the field, though WKB corrections to the quasinormal modes of order of $1/\ell$ and higher may be quantitatively different for different fields.

As an illustration we shall consider first the metric which has $b_i = \Phi_i =0$ for $i\geq 2$. Substituting the Taylor expanded form of the functions (\ref{b-expansion}, \ref{ph-expansion}) into the formula (\ref{WKB}) at the second WKB order ($p=2$)   and expanding the result in powers of $1/\ell$ we obtain the analytical expression for the quasinormal modes:
$$ \omega = \frac{e^{\Phi _0}}{b_0} \left(\ell + \frac{1}{2}\right) - \imo\frac{e^{\Phi_{0}}}{\sqrt{2} b_{0}} \left(n+\frac{1}{2}\right)
\sqrt{(b_{1} -1)(b_0 \Phi_{1} -1)}  ~ +  $$
\begin{equation}\label{Omega}
\frac{e^{\Phi _0} \left(-6 b_0^2 \left(b_1-1\right) \Phi _1^2+2 b_0 \left(5 b_1-9\right) \Phi _1-7 b_1+15\right)}{64 b_0 \left(b_0 \Phi _1-1\right)\ell} +  {\cal O}(1/\ell^2).
\end{equation}
Notice that in order to obtain accurate quasinormal modes up to $\sim 1/\ell$ order, the first order WKB is insufficient.

Let us now suppose that we know the fundamental quasinormal modes for several large multipole numbers $\ell$. Then we can fit the numerical data obtained from quasinormal modes at different $\ell$ by the following general formula
\begin{equation}
\omega(\ell) = A \left(\ell + \frac{1}{2}\right)  - B \imo  + C \ell^{-1},
\end{equation}
where the coefficients $A$, $B$ and $C$ are related to the coefficients of the Taylor expansion of the metric functions as follows
\begin{equation}\label{AB}
A = \frac{e^{\Phi _0}}{b_0}, \quad \frac{B}{e^{\Phi _0}}= \frac{\sqrt{\left(b_1-1\right) \left(b_0 \Phi _1-1\right)}}{2 \sqrt{2} b_0},
\end{equation}
\begin{equation}\label{C}
\frac{C}{e^{\Phi _0} }= \frac{6 b_0^2 \left(1- b_1 \right) \Phi _1^2+2 b_0 \left(5 b_1-9\right) \Phi _1-7 b_1+15}{64 b_0 \left(b_0 \Phi _1-1\right)}.
\end{equation}

Solving these three algebraic equations (\ref{AB}, \ref{C}) relatively $\Phi_{0}$, $b_{1}$ and $\Phi_{1}$ we can define the first derivatives of the shape and redshift function at the throat and values of the redshift function at the throat.
This gives us the wormhole geometry near the throat up to the first order.
\begin{table}\label{Table1}
  \centering
  \begin{tabular}{|c|c|c|c|}
     \hline
     $\ell$ & $\Phi _0$ & $ \Phi _1$  & $ b _1$  \\
       \hline
     100-110 & 2.9999997 & -1.91279 & -5.4416 \\
         \hline
     1000-1010 & 2.9999999 & -1.98987 & -5.0489 \\
          \hline
     10000-10010 & 2.9999999 & -1.99906 & -5.0045 \\
          \hline
     exact & 3 & -2 & -5 \\
     \hline
   \end{tabular}
  \caption{Fitting of quasinormal modes for electromagnetic perturbations at high $\ell$: approximate values of the coefficients $b_1$, $\Phi _0$ and $ \Phi _1$  versus their exact values.} \label{table1}
\end{table}
From Table \ref{table1}, one can see that at $\ell \sim 10^3$  the relative error of the fitting for $b_1$ and $\Phi_1$ is less than one percent.

If the wormhole is tideless, then $\Phi(r) = \Phi_{0}$, so that all the derivatives of $\Phi$ vanish and relation (\ref{Omega}) takes the simple form:
\begin{equation}\label{A1}
\omega = \frac{e^{\Phi _0}}{b_0} \left(\ell + \frac{1}{2}\right) - \imo\frac{e^{\Phi_{0}}}{\sqrt{2} b_{0}} \left(n+\frac{1}{2}\right)
\sqrt{(1- b_{1})}.
\end{equation}
This way, the parameters of the near throat geometry $\Phi_{0}$ and $b_{1}$ can be immediately read from the real oscillation frequency and the damping rates of the modes as follows:
\begin{equation}\label{A2}
\Phi_{0} = \log\left( \frac{b_{0} Re (\omega)}{\ell+\frac{1}{2}}\right), \quad b_{1} = 1- \sqrt{-\frac{2 (\ell+ \frac{1}{2})}{b_{0} (n+ \frac{1}{2})}  \frac{Im (\omega)}{Re (\omega)}.  }
\end{equation}
Here we have considered the example of wormhole whose metric functions $\Phi(r)$ and $b(r)$ are linear in $r$. If one assumes the general Taylor expansion for both metric functions, no such simple fitting is possible in the general case.

\section{Higher order terms}

Now, let us consider the general metric given by its Taylor series (\ref{b-expansion}, \ref{ph-expansion}) where we shall keep terms up to the third order (thus, $a_i = \Phi_i =0$ for $i \geq 4$). After some calculations it can be seen that higher than the third order terms in the Taylor expansion do not influence the WKB series up to the $1/\ell^2$ order. In order to make all the equations simpler, from now and on let us use the new parameter $k=1/(\ell(\ell+1))$ instead of $\ell$. In the high $k$ regime we have
\begin{equation}\label{highLgen}
\omega_n = \frac{e^{\Phi _0}}{b_0 \sqrt{k}}-\frac{i e^{\Phi _0} \sqrt{\left(b_1-1\right) \left(b_0 \Phi _1-1\right)}}{\sqrt{2} b_0} \left(n+\frac{1}{2}\right) + C_{n} \sqrt{k} + {\cal O}(k),
\end{equation}
where for $n=0$
\begin{equation}
C_{0}= \frac{\sqrt{k} e^{\Phi _0} \left(b_0^2 \left(6 \left(\Phi _1^2+\Phi _2\right)-2 b_2 \Phi _1\right)+2 b_0 \left(b_2-5 \Phi _1\right)-b_1 \left(6 b_0^2 \left(\Phi
   _1^2+\Phi _2\right)-10 b_0 \Phi _1+7\right)+7\right)}{64 b_0 \left(b_0 \Phi _1-1\right)}.
\end{equation}

Notice that the first two terms in (\ref{highLgen}) do not depend on the terms which are higher than the first order of the metric Taylor expansion. Accurate computation of the quasinormal modes must lead the following general formula
  \begin{equation}\label{highLgen2}
\omega_n = \frac{A}{\sqrt{k}}- B\left(n+\frac{1}{2}\right) + C_{n} \sqrt{k} + {\cal O}(k).
\end{equation}
The above equation will give us three constants $A$, $B$ and $C_n$, for each $n$. This cannot be used to find five coefficients $b_1$, $b_2$ $\Phi_0$, $\Phi_1$, $\Phi_2$ through the fitting of the corresponding algebraic equations. Thus, only provided the redshift function is known independently, i. e. $\Phi_i$ are determined, then one can find the Taylor expansion for the shape function of the wormhole $b_1$, $b_2$, etc., consequently in each term of the expansion in powers of $k$. In the particular case, if the wormhole is known to be tideless, as in the example considered below, the procedure is simpler and allows us to find higher order terms of the Taylor expansion for the shape function.

Notice, that the usage of several equations for various overtones $n$ will not remedy the situation, as these additional equations will be proportional to the one for the fundamental mode and thus will not provide any algebraically independent equations, which are necessary for fitting.
Though, in a similar fashion with \cite{Volkel:2017kfj},\cite{Volkel:2018hwb},  the inversion of the Bohr-Sommerfeld rule might produce different set of equations which would be independent for different $n$. This does not happen when one directly applies the Will-Schutz-Iyer formula of \cite{WKB,WKBorder}.

\section{Examples}

\subsection{Bronnikov-Ellis wormhole}

Suppose that we are limited by the tideless wormholes for which all $\Phi_i =0$.
Then, the third order WKB formula for $\omega$ gives the following expansion in terms of $k=1/(\ell(\ell+1))$:

$$ \omega = \frac{1}{b_0 \sqrt{k}} -\frac{i \sqrt{1-b_1}}{2 \sqrt{2} b_0} - \frac{\left(-7 b_1+2 b_0 b_2+7\right) \sqrt{k}}{64 b_0}+$$
\begin{equation}\label{qnm-gen}
\frac{i \left(-4 \left(b_2^2+6 b_3\right) b_0^2+36 b_2 b_0+51 b_1^2+6 b_1 \left(4 b_3 b_0^2-6 b_2
   b_0-17\right)+51\right) k}{1024 b_0 \sqrt{2-2 b_1}}+ {\cal O}(k^{3/2})
\end{equation}

The Bronnikov-Ellis wormhole has the following metric functions \cite{Bronnikov:1973fh}:
\begin{equation}
\Phi(r) = 0, \quad b(r) = \frac{b_{0}^2}{r}.
\end{equation}
We can use the above formulas for the 6th order WKB formula or any other approach (for example the Frobenius method) for getting accurate quasinormal modes in the regime of high $\ell$. The results can be fitted with by the following formula:
\begin{equation}\label{qnm-1}
\omega =  \frac{1}{b_0 \sqrt{k}}-\frac{i}{2 b_0} -\frac{\sqrt{k}}{4 b_0} + \frac{5 i k}{32 b_0} + {\cal O}(k^{3/2}).
\end{equation}
Choosing the radius of the throat $b_0 =1$ and equating the terms at the same orders of $k$ in eqs. (\ref{qnm-gen}) and  (\ref{qnm-1}), we find that
\begin{equation}
b_1= -1, \quad b_2 =1, \quad b_3= -1.
\end{equation}
Thus, we reproduce the series expansion of $1/r$ in terms of  $(r-1)$, which is
\begin{equation}
b(r) = \frac{1}{r} = 1 - (r-1) + (r-1)^2 - (r-1)^3 + {\cal O}((r-1)^{4}).
\end{equation}

The same procedure can be used for reconstruction of the shape function of an arbitrary traversable tideless wormhole near its throat.

\subsection{Tideless Morris-Thorne wormhole}

Here we shall consider the tideless Morris-Thorne wormhole suggested in eqs. (A1) of \cite{Morris-Thorne}.
The shape function has the form:
\begin{equation}
b(r) = \sqrt{b_{0} r}.
\end{equation}
We find that the for the above metric the high $\ell$ quasinormal frequencies obey the following asymptotic:
\begin{equation}\label{qnm-2}
\omega =  \frac{1}{b_0 \sqrt{k}}-\frac{i}{4 b_0} -\frac{13 \sqrt{k}}{256 b_0} +\frac{155 i k}{16384 b_0} +  {\cal O}(k^{3/2}).
\end{equation}
One can easily see that taking $b_{0} =1$ and comparing the terms at the same powers of $k$ in eqs. (\ref{qnm-gen}) and  (\ref{qnm-2}), we find that
\begin{equation}
b_{1}= \frac{1}{2}, \quad b_{2}= -\frac{1}{8}, \quad b_{3}= \frac{1}{16},
\end{equation}
which are the coefficients of the Taylor series for $\sqrt{r}$.

\section{Final remarks}

Here we have shown that the shape function of a spherically symmetric traversable Lorentzian wormhole near its throat can be reconstructed from high frequency quasinormal modes of the wormhole. To the best of our knowledge this is the first, although not the most general, solution of the inverse problem for a compact object, which leads from quasinormal modes not to a family of effective potentials, but directly to the metric. The near throat geometry is observationally much more important for a wormhole, than the near horizon geometry of a black hole, because the peak of the effective potential is located at the throat, so that scattering of fields, accretion and other astrophysically relevant phenomena happen near the wormhole throat.

We have shown that the unique solution to the inverse problem  is possible within the WKB approach for tideless wormholes or once the tidal force is known independently, for instance, through the redshift effect from various sources. In the general case, WKB approach, which we used here, does not lead to the unique solution for the metric.

It seems that the solution to the inverse problem presented in this paper, as well as other approaches to the inverse problem in the literature (see e.g. \cite{Volkel:2017kfj,Volkel:2018hwb} and references therein) cannot be applied for interpretation of LIGO/VIRGO or other new future experiments. This happens because the outcome of such an experiment is only one mode per event and, provided we observe many events, possibly a few modes corresponding to a few relatively low multipole numbers. This is evidently not sufficient to provide a qualitative fit for extracting the wormhole geometry.

Posing the throat at the peak of the effective potential does not effect the possibility of a wormhole to mimic the black hole ring-down. This freedom to mimic black holes comes from the two arbitrary functions, the shape function and the red-shift function. However, as it was shown in \cite{Konoplya:2016hmd} wormholes which are non-symmetric relatively its throat are more effective mimickers of black holes and can produce many modes of the spectrum of a black hole.

Our paper can be extended in a number of ways. First of all, in order to avoid lengthy formulas, we considered the electromagnetic field only. Other test fields can also be studied within the same approach. Then, the higher order Taylor expansion of the metric requires usage of higher WKB orders in order to get accurate values of quasinormal modes up to higher orders of $1/\ell$. Some of our results, for example formulas (\ref{A1},\ref{A2}) are valid for a test scalar field as well. Finally, our approach could be applied also to rotating wormholes provided they are symmetric enough to guarantee the separation of variables with the know spherical harmonics. In this case, the fitting could be done via the WKB approach.

\acknowledgments{
The author was supported by the Mobility Grant at Silesian University in Opava. The publication has been prepared with the support of the ``RUDN University Program 5-100''. The author would also like to thank Alexander Zhidenko, Sebastian V\"{o}lkel, Kostas Kokkotas and Kirill Bronnikov for useful discussions.}

\end{document}